# Machine learning-assisted discovery of many new solid Li-ion conducting materials

Austin D. Sendek,[1] Ekin D. Cubuk,[2,3] Evan R. Antoniuk,[4] Gowoon Cheon,[1] Yi Cui,[2] Evan J. Reed[2]*

1. Department of Applied Physics, Stanford University, Stanford, CA, USA
2. Department of Materials Science and Engineering, Stanford University, Stanford, CA, USA
3. Google Brain, Mountain View, CA, USA
4. Department of Chemistry, Stanford University, Stanford, CA, USA

*Corresponding author: evanreed@stanford.edu

**Abstract**

We discover many new crystalline solid materials with fast single crystal Li ion conductivity at room temperature, discovered through density functional theory simulations guided by machine learning-based methods. The discovery of new solid Li superionic conductors is of critical importance to the development of safe all-solid-state Li-ion batteries. With a predictive universal structure-property relationship for fast ion conduction not well understood, the search for new solid Li ion conductors has relied largely on trial-and-error computational and experimental searches over the last several decades. In this work, we perform a guided search of materials space with a machine learning (ML)-based prediction model for material selection and density functional theory molecular dynamics (DFT-MD) simulations for calculating ionic conductivity. These materials are screened from over 12,000 experimentally synthesized and characterized candidates with very diverse structures and compositions. When compared to a random search of materials space, we find that the ML-guided search is 2.7 times more likely to identify fast Li ion conductors, with at least a 45x improvement in the log-average of room temperature Li ion conductivity. The F1 score of the ML-based model is 0.50, 3.5 times better than the F1 score expected from completely random guesswork. In a head-to-head competition against six Ph.D. students working in the field, we find that the ML-based model doubles the F1 score of human experts in its ability to identify fast Li-ion conductors from atomistic structure with a thousand-fold increase in speed, clearly demonstrating the utility of this model for the research community. In addition to having high predicted Li-ion conductivity, all materials reported here lack transition metals to enhance stability against reduction by Li metal anode, and are predicted to exhibit low electronic conduction, high stability against oxidation, and high thermodynamic stability, making them promising candidates for solid-state electrolyte applications on these several essential metrics.

## I. Introduction

All-solid-state Li-ion batteries (SSLIBs) hold promise as safer, longer lasting, and higher energy dense alternatives to today's commercialized LIBs with liquid electrolytes. However, the design of SSLIBs remains a challenge, with the principal technological bottleneck in realizing these devices being the solid electrolyte. A high performance solid electrolyte material must satisfy several criteria simultaneously: it must possess fast Li-ion conduction, negligible electronic conduction, a wide electrochemical window, robust chemical stability against side reactions with the electrodes, and high mechanical rigidity to suppress dendritic growth on the anode. The material should also be cheap and easy to manufacture. Given these many constraints, searching for promising new materials that satisfy all requirements through the trial-and-error searches has yielded slow progress.

The earliest efforts to discover fast Li-ion conducting solids began in the 1970s[1] and have continued to present. More recently, density functional theory (DFT) simulation has enabled high-throughput computational searches, essentially automating the process of guess-and-check.[2,3] Across these four decades, only several solids with liquid-level Li conduction (>$10^{-2}$ S/cm) at room temperature (RT) have been identified, including notably $Li_{10}GeP_2S_{12}$ (17 mS/cm),[4] $Li_7P_3S_{11}$ (25 mS/cm).[5] This slow progress suggests that continuing in the guess-and-check paradigm of decades past is unlikely to quickly yield the material innovations we need to unlock the high energy density, high cycle life, and unquestionably safe energy storage devices of the future.

Leveraging atomic and electronic structure data from the Materials Project database,[6] we have screened all 12,000+ Li-containing materials for thermodynamic phase stability, low electronic conduction, high electrochemical stability, and no transition metals (to enhance stability against reduction). We also compiled information on the estimated raw materials cost and the earth abundance of the elemental constituents of each material. This identifies 317 materials that may be strong candidate electrolyte materials if they are also fast ion conductors.

Following the guess-and-check paradigm, one would begin to test these materials for fast ion conduction at random, or according to his/her best scientific intuition. To identify the subset of these materials most likely to exhibit fast ionic conductivity, we have instead developed[7] a machine learning (ML)-based model for predicting the likelihood $P_{superionic}$ that an arbitrary material exhibits fast Li-ion conduction at RT, based only on features $x_i$ derived from the atomistic structure of the unit cell. Throughout this work, we define superionic conductivity to be greater than 0.1 mS/cm, based on the approximate minimum electrolyte ionic conductivity required for battery applications. Experimental reports of ionic conductivity for several dozen materials ranging over 10 orders of magnitude were used to train the model. This data-driven predictor takes the form of a logistic function, $P_{superionic}(x) = (1 + e^{-\Sigma_i \Theta_i x_i})^{-1}$, where:

$$\sum_i \Theta_i x_i = 0.18 \times LLB - 4.01 \times SBI - 0.47 \times AFC + 8.70 \times LASD - 2.17 \times LLSD - 6.56. \tag{1}$$

Here, *LLB* is the average Li-Li bond number (number of Li neighbors) per Li; *SBI* is the average sublattice bond ionicity; *AFC* is the average coordination of the anions in the anion framework; *LASD* is the average equilibrium Li-anion separation distance; and *LLSD* is the average equilibrium Li-Li separation distance. Since this model does not require any electronic structure information, it is >5 orders of magnitude faster to evaluate than a DFT simulation of conductivity.

Screening this list of 317 candidate materials identifies 21 crystalline compounds that are predicted to be fast ion conductors with robust structural and electrochemical stability, representing a 99.8% reduction in the entire space of known Li-containing materials. One of these 21 materials, LiCl, has been reported to exhibit poor RT Li conduction (~$10^{-9}$ S/cm),[8] making it a known false positive prediction. Another material, $Li_3InCl_6$, has been reported to have a RT conductivity of approximately 1 mS/cm, making it a correct model prediction.[9] Very little is reported in the literature regarding the remaining 19 materials to our knowledge.

In this work, we perform DFT molecular dynamics (DFT-MD) calculations[10,11] on the promising candidate materials identified by our screening procedure, finding evidence of superionic RT Li conduction in eight and marginal RT Li conduction in two.† As a control, we then perform DFT-MD on a similar number of materials drawn at random from the same population of 317. We quantify the increase in research efficiency offered by our ML-based model by comparing the improvements in experimental outcomes against the random case. We consider 41 unique materials in total. We find our ML-guided search offers approximately 3-4x improvement in predictive power for fast Li ion conductors over random guesswork depending on the metric, while on average the predicted RT Li ionic conductivity of any simulated material is over 45x higher.

As a further test of the model's efficacy, we provided the same list of materials to a group of six graduate students working in the field and asked them to identify the best ion conductors. We found the F1 score of the model outperformed the F1 score of the intuition of the students by approximately two times, while each prediction was made approximately 1,000 times faster. This result suggests ML-based approaches to materials selection may provide significant acceleration over the guess-and-check research paradigm of the past. Furthermore, these results provide confidence in our data-driven superionic prediction model, as well as compelling evidence in the promise of machine learning-based approaches to materials discovery.

**II. Simulations**

We first perform DFT-MD on the 19 most promising new candidate materials for solid electrolyte applications that are all predicted to be fast ion conductors by our ML-based model Eq. (1). LiCl and $Li_3InCl_6$ were not simulated due to the existence of conductivity data in the literature.[8,9] In order to accelerate Li diffusion to a computationally tractable timescale, we initially seek an upper bound by performing MD at elevated temperature and removing one Li atom per computational cell to introduce a small concentration of Li vacancies to enhance conduction and minimize the number of false negatives identified. We simulate large supercells in order to minimize the effect of the periodic boundary conditions. All materials were initially simulated at $T = 900$ K; if melting is observed, the simulation is restarted at increasingly lower temperatures until no melting is observed. The vacancy concentration ranges from 3-17% depending on the unit cell. The simulation temperatures, computational cell size, and Li vacancy concentrations are provided in the Supporting Information, Table S1.

We simulate the candidate materials for a range of times on the tens to hundreds of picoseconds timescale; see Table S1. To calculate ionic diffusivity, which may be isotropic, we compute the average of the diagonal elements of the 2$^{nd}$ rank Li diffusivity tensor (or equivalently, one-third of the trace). We denote this as $\langle D_{ii} \rangle$, where the average is taken over the three elements $ii = \{xx, yy, zz\}$. We first evaluate the mean squared displacement (MSD) of the Li atoms $\langle \Delta r^2 \rangle$ over time (starting at $t = 0$) and apply the following formula:

$$\langle D_{ii} \rangle = \lim_{t \to \infty} \frac{1}{6} \frac{\langle \Delta r^2 \rangle}{\Delta t} \qquad (2)$$

To probe for melting, we also calculate the MSD of the sublattice atoms and assume melting if sustained, non-zero diffusivity is observed in both Li and the sublattice (the smallest diffusivity that can be resolved through DFT-MD at 900K is approximately 0.01 Å$^2$/ps). To assess the degree of convergence in $\langle D_{ii} \rangle$, we compute the standard deviation in diffusivities when measured from different starting times in the MSD data. We compute the slope of the MSD data for every starting time from t = 0 to up to 75% of the total run time, in 100 fs increments. We compute one standard deviation above and below the mean diffusivity across all starting times to represent the approximate upper and lower limits of the distribution of diffusivities one may observe under these simulation conditions. The diffusivity as measured from $t = 0$ and the mean diffusivity across all starting times are not necessarily equivalent, and thus the upper and lower uncertainties are not necessarily symmetric around the value predicted from $t = 0$.

Ionic transport in crystalline solids is modeled here as a stochastic phenomenon governed by a Boltzmann (Arrhenius) factor that is exponential in a single energy barrier between equilibrium sites, $E_a$:

$$\langle D_{ii} \rangle(T) = D_0 e^{-E_a/k_B T} \tag{3}$$

Ionic hopping becomes exponentially less likely as the energy barrier increases, and at 900 K ($k_B T$ = 78 meV) hopping may not happen on the tens to hundreds of picoseconds timescale if the barrier is much above 0.4 eV ($e^{-0.4/0.078} = 0.006$). Thus, high temperature MD simulations are not well-suited to predict a numerical value for ionic conductivity in medium- to high barrier systems because the statistical behavior is not captured on typical DFT simulation timescales. Numerical values can only be predicted through MD in low barrier systems, and even then are still approximate due to the stochastic nature of hopping. Our approach is most suited for high throughput calculations in which the main goal is identifying whether or not materials are superionic conductors, i.e. for making binary predictions of superionic vs. non-superionic conduction.

From the diffusivity $\langle D_{ii} \rangle$, we convert to the average of the diagonal elements of the ionic conductivity tensor $\langle \sigma_{ii} \rangle$ via the Einstein relation:[12]

$$\langle \sigma_{ii} \rangle(T) = \frac{\langle D_{ii} \rangle(T) n q^2}{k_{\mathrm{B}} T} \tag{4}$$

where $n$ is the concentration of Li atoms and $q$ is the charge on the Li atoms. We measure the effective Li charge $q$ using the Bader charge analysis methods of Henkelman et al.[13–16] Given the linear relationship between conductivity and diffusivity, we compute the spread of possible conductivities by applying Eq. (4) to the computed bounds in diffusivity as well.

It stands to reason that fast Li conduction will not be observed at RT if it is not observed in these favorable conditions. Our approach therefore should not identify false negatives (materials with poor conduction at high temperature but fast conduction at RT), although there is a risk of identifying false positives (materials with fast conduction at high temperature but poor conduction at RT). To ensure favorable scaling to room temperature and guard against the identification of false positives, we simulate DFT-MD again at alternate temperatures if significant Li diffusion has been observed at the initial simulation temperature. The line connecting the two or three diffusivities on an Arrhenius plot of $\log_{10}(\sigma)$ versus inverse temperature is employed here to extrapolate down to RT without direct calculation of $E_{\mathrm{a}}$. This assumes Arrhenius scaling applies and there are no structural phase changes or non-structural superionic transitions (where conductivity changes abruptly as new ionic pathways become energetically accessible or inaccessible without any significant sublattice rearrangement)[17] between the simulation temperatures and RT. The spread of possible values in the RT conductivity extrapolations is computed by extrapolating along the upper and lower limits of the deviations in the high temperature conductivity calculations. Using typical values for Li concentration, Bader charge, and site hopping distance, we estimate that the RT conductivity is likely to be $10^{-9}$ S/cm or lower if no Li diffusion is observed on the simulation timescale at 900K (see Supporting Information, Section S1 for calculation).

A flowchart describing this process is shown in Fig. 1. This extrapolation scheme requires less computational expense than simulating each material at several different temperatures right away, and is the most tractable way for us to quantify the ionic conductivity of the materials studied here on a reasonable time scale. Screening for Li superionic RT diffusion by two-temperature extrapolation has been leveraged in recent work to search sulfide-based compositional spaces, for example.[3]

We perform DFT calculations with the Vienna Ab Initio Simulation Package (VASP)[18] with the generalized gradient approximation (GGA) of Perdew-Burke-Ernzerhof (PBE)[19] and the projector augmented wave (PAW)[20] method. We use the pseudopotentials and plane wave cutoff energy (520 eV for all structures) as recommended by the Materials Project. The VASP input files are generated using the *pymatgen.io.vasp.sets* module of Pymatgen.[21,22] Given the large unit cells (and DFT Kohn-Sham bandgaps exceeding 1 eV for all these materials), we use a gamma-point only *k*-mesh. The pseudopotentials are given in Table S1.

The ultimate metric of the utility of materials selection models is the improvement in likelihood of successfully identifying positive examples over the background probability of these materials. To quantify the superiority of our model to completely random guesswork, we must know the likelihood of discovering a superionic material by chance with no scientific intuition involved. In 1978, Huggins and Rabenau questioned "whether there is anything fundamentally different in such materials, or whether they merely exhibit extreme values of 'normal' behavior."[23] Superionic conductors may have remarkably low diffusion barriers, but how do they compare to the distribution of barriers across the space of all the candidate electrolyte materials?

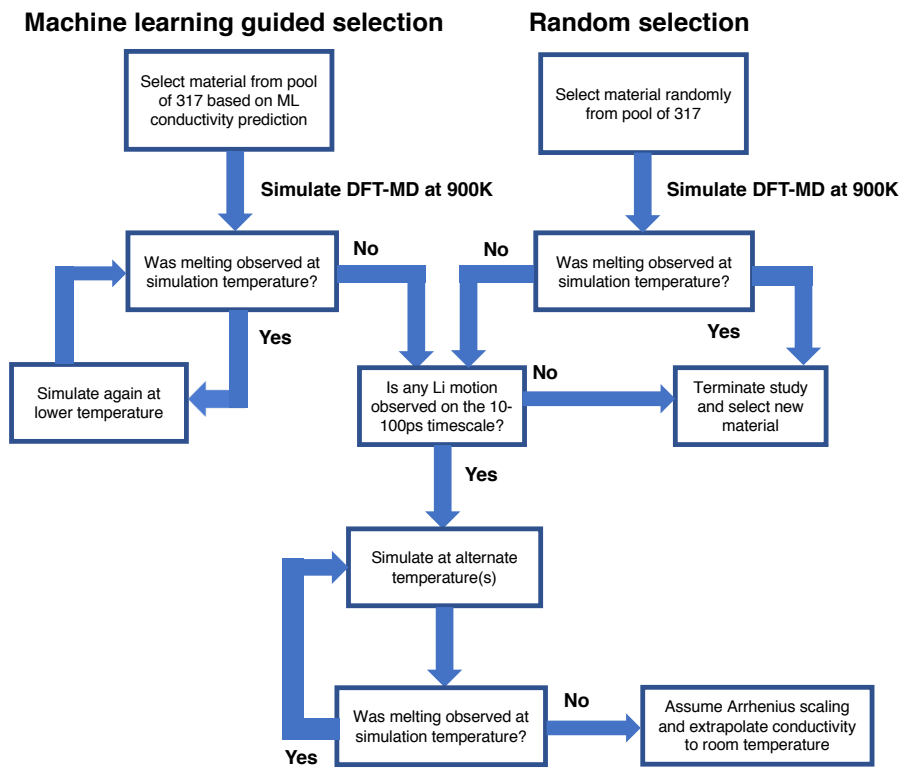

**Figure 1:** *Flowchart of high-throughput density functional theory molecular dynamics (DFT-MD) simulation process.* Materials are chosen either according to the machine learning-based model (left stream) or by random selection (right stream). Materials are initially simulated at 900K for approximately 50-100 picoseconds. If no Li diffusion is observed on this timescale, the materials are immediately considered poor ion conductors and no further simulation is performed. If materials exhibit Li diffusion at 900K, they are simulated again at alternate temperatures so a simple two- or three-point Arrhenius extrapolation to room temperature can be made. If melting is observed in the ML-selected materials at 900K, the simulation is restarted at a lower temperature. Randomly chosen materials that exhibit melting at 900K are discarded immediately for computational efficiency; this occurred in two randomly chosen materials.

Experience would tell us that the likelihood of chance discoveries is low, given that only a handful of fast Li conductors have been discovered since the search began nearly 40 years ago. However, we are probing for superionic conduction in a very specific way: we are pulling from a potentially biased group of 317 known materials in the Materials Project database, introducing Li vacancies, and simulating finite computational cells at high temperature. Therefore, this question is best answered by doing a straight-across comparison of high temperature DFT-MD simulations for a similar number of structures chosen at random from the same population.

To accomplish this, we perform a control experiment where we simulate 21 structures chosen uniformly at random from among the 317 materials that satisfy all prerequisite screening criteria (band gap > 1 eV, predicted oxidation potential > 4 V, energy above convex hull = 0 eV, no transition metals) and simulate MD under the same procedure. These 21 structures and their predicted superionic likelihoods according to the ML model are the following: $LiLa_2SbO_6$ (0%), $Li_6UO_6$ (7.7%), $LiInF_4$ (0.6%), $LiBiF_4$ (0.2%), $Li_6Ho(BO_3)_3$ (10.1%), $RbLiB_4O_7$ (4%), $Li_4Be_3As_3ClO_{12}$ (0.3%), $Li_6TeO_6$ (6.4%), $Li_3Pr_2(BO_3)_3$ (9.2%), NaLiS (36%), $LiSbO_3$ (2.4%), $LiCaGaF_6$ (0.0%), $Li_2Te_2O_5$ (15.8%), $LiNO_3$ (8.3%), $Ba_4Li(SbO_4)_3$ (0.0%), $Rb_2Li_2SiO_4$ (4.5%), $NaLi_2PO_4$ (0.0%), $Cs_4Li_2(Si_2O_5)_3$ (0.1%), $RbLi(H_2N)_2$ (24.5%), $Cs_2LiTlF_6$ (0.0%), and $LiSO_3F$

(100%). We chose 29 structures in total but removed eight ($Na_2LiNF_6$, $CsLi_2(HO)_3$, $LiU_4P_3O_{20}$, $Li_3Nd_2H_6(N_3O_{10})_3$, $Li_3La_2H_6(N_3O_{10})_3$, $Li_3P_{11}(H_3N)_{17}$, $Li_4H_3BrO_3$, and $Li_2Pr(NO_3)_5$) due to either melting at 900K or due to extremely slow or failed electronic convergence, in some cases due to electronic bandgap closure during the simulation. Of the remaining 21 randomly chosen materials we successfully simulate, 20 have a negative ML-based superionic prediction (below <50%) so we do not expect them to conduct. The only material of these 21 random materials that is predicted by our ML-based algorithm to conduct is $LiSO_3F$. The computational parameters used in these randomly chosen simulations are also provided in Table S1. This set of randomly-drawn materials also provides a test set to explore the ML model performance versus the predictive power of the intuition of human experts; we explore this in section III, subsection v.

In this work we report a total of 4.3 nanoseconds of molecular dynamics simulation, with a mean simulation time of 74.9 picoseconds and mean computational cell size of 99.1 atoms. The total volume of data reported here corresponds to approximately 330,000 GPU-hours of simulation.

## III. Results and Discussion

*i. Discoveries from ML-guided selection*

The computed ionic conductivities for the ML-chosen candidate materials and the randomly chosen materials are provided in the Arrhenius diagram on Fig. 2(a). The computed high temperature diffusivities, ionic conductivities, average Li Bader charge and extrapolated RT ionic conductivities with predicted deviations (conductivities measured from alternate starting time points) of these candidates are given in Table 1. We do not calculate results for one of the

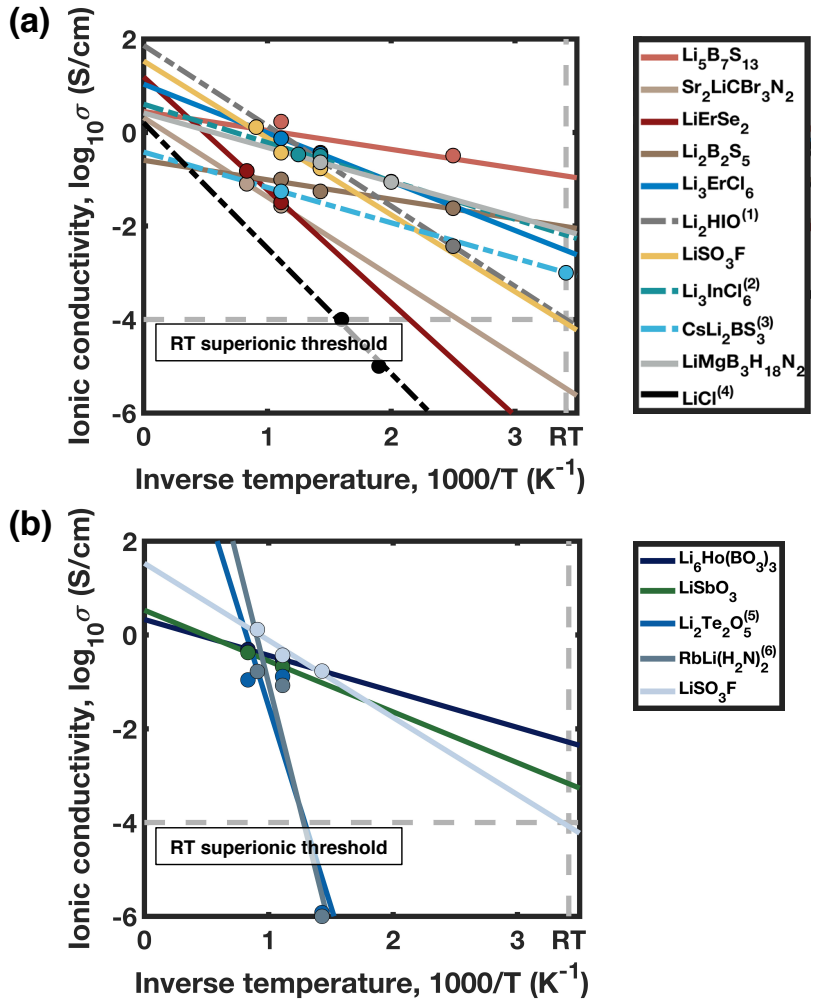

**Figure 2:** *Computationally observed ionic conductivity in candidate materials and extrapolation to room temperature (RT).* In fig. (a) we show the computed ionic conductivities for the ten fast conducting materials as chosen by the ML model (from 20 of the 21 identified in [7]; $Li_2GePbS_4$ not simulated); in fig. (b) we show the same for the five conducting materials from the 21 randomly chosen materials. The vertical dotted grey line represents room temperature (293K) and the horizontal dotted grey line represents our RT ionic conductivity threshold ($10^{-4}$ S/cm) for a material to be considered superionic. All calculations were performed with small Li vacancy concentrations as listed in Table S1 in the Supporting Information. Materials simulated in this work are not shown here if no Li conduction was observed. After extrapolating to room temperature, ten of the ML-chosen materials exhibit significant RT ion conductivity (with eight of them at or above $10^{-4}$ S/cm) in (a), while the same is true of only three of the randomly chosen materials in (b). The predicted RT conductivity values are provided in Tables 1 and 2. The errors associated with these calculations are not shown here but are listed in Tables 1 and 2. Materials with dotted lines have caveats in their extrapolations: (1) $Li_2HIO$ was only simulated at 400K due to melting at higher temperatures; the RT extrapolation assumes a diffusion barrier of 0.35 eV, a typical, if high, Li ion diffusion barrier in fast conducting materials. (2) $Li_3InCl_6$ was not simulated in this work and conductivity values were taken from [9]. (3) $CsLi_2BS_3$ was initially simulated at 900K but issues with slow or failed convergence were encountered when simulating at alternate temperatures. The RT conductivity value is assumed to be of the same order of magnitude as $Li_3BS_3$ in [24]. (4) LiCl was also not simulated here with conductivity values taken from [8]. Both $Li_2Te_2O_5$ (5) and $RbLi(H_2N)_2$ (6) did not exhibit any Li conduction at 700K on the 50ps timescale so a conductivity $\leq 10^{-6}$ S/cm at this temperature was assumed.

| MPID (mp-####) | Chemical formula | Computed $\langle D_{ii} \rangle$ ($Å^2$/ps), $T$ (K) | Average Li Bader valence charge | Computed $\langle \sigma_{ii} \rangle$ (mS/cm), T (K) | Extrapolated RT $\langle \sigma_{ii} \rangle$ (mS/cm) | ML-predicted superionic? | Evidence for correct model prediction? |
|---|---|---|---|---|---|---|---|
| 554076 | BaLiBS$_3$ | ~0 [**900**] | – | ~0 [**900**] | < $10^{-6}$ | Yes | No |
| 532413 | Li$_5$B$_7$S$_{13}$ | 0.88 (1.0, 0.75) [**900**], 0.16 (0.23, 0.094) [**700**], 0.078 (0.090, 0.052) [**400**] | +0.89 | 1700 (1800, 1600) [**900**], 370 (540, 220) [**700**], 320 (370, 220) [**400**] | 120 (140, 67) | Yes | Yes |
| 569782 | Sr$_2$LiCBr$_3$N$_2$ | 0.15 (0.15, 0.047) [**1200**], 0.038 (0.039, 0.027) [**900**] | +0.90 | 80 (84, 26) [**1200**], 27 (28, 20) [**900**] | 0.0033 (0.0031, 2.3) | Yes | Marginal |
| 558219 | SrLi(BS$_2$)$_3$ | ~0 [**900**] | – | ~0 [**900**] | < $10^{-6}$ | Yes | No |
| 15797 | LiErSe$_2$ | 0.12 (0.16, 0.13) [**1200**], 0.020 (0.024, 0) [**900**] | +0.89 | 150 (190, 160) [**1200**], 32 (40, 0) [**900**] | $8.8 \times 10^{-5}$ ($9.9 \times 10^{-5}$, 0) | Yes | Marginal |
| 29410 | Li$_2$B$_2$S$_5$ | 0.058 (0.082, 0.058) [**900**], 0.025 (0.03, 0.015) [**700**], 0.0063 (0.0046, 0) [**400**] | +0.90 | 99 (140, 99) [**900**], 55 (65, 32) [**700**], 24 (17, 0) [**400**] | 9.7 (4.4, 0) | Yes | Yes |
| 676361 | Li$_3$ErCl$_6$ | 0.34 (0.40, 0.24) [**900**], 0.12 (0.11, 0.037) [**700**] | +0.91 | 750 (870, 530) [**900**], 350 (300, 100) [**700**] | 3 (0.39, 0.10) | Yes | Yes |
| 643069 | Li$_2$HIO | 0.00053 (0.000048, 0) [**400**] | +0.88 | 3.7 (0.34, 0) [**400**] | ~0.1* | Yes | Yes |
| 7744 | LiSO$_3$F | 0.86 (0.92, 0.65) [**1100**], 0.20 (0.22, 0.17) [**900**], 0.072 (0.08, 0.059) [**700**] | +0.91 | 1300 (1400, 960) [**1100**], 370 (390, 310) [**900**], 170 (190, 140) [**700**] | 0.082 (0.11, 0.10) | Yes | Yes |
| 34477 | LiSmS$_2$ | ~0 [**900**] | – | ~0 [**900**] | < $10^{-6}$ | Yes | No |
| 676109 | Li$_3$InCl$_6$ | N/S | N/S | N/S | 6.4 [from Ref. [9]] | Yes | Yes |
| 559238 | CsLi$_2$BS$_3$ | 0.055 (0.11, 0.066) [**900**] | +0.87 | 100 (190, 120) [**900**] | ~1 [from Ref. [24]]† | Yes | Yes |
| 866665 | LiMgB$_3$(H$_9$N)$_2$ | 0.26 (0.78, 0.29) [**700**], 0.07 (0.17, 0.08) [**500**] | +0.90 | 230 (690, 260) [**700**], 87 (210, 100) [**500**] | 7.9 (11, 9.4) | Yes | Yes |
| 8751 | RbLiS | ~0 [**900**] | – | ~0 [**900**] | < $10^{-6}$ | Yes | No |
| 15789 | LiDyS$_2$ | ~0 [**900**] | – | ~0 [**900**] | < $10^{-6}$ | Yes | No |
| 15790 | LiHoS$_2$ | ~0 [**900**] | – | ~0 [**900**] | < $10^{-6}$ | Yes | No |
| 15791 | LiErS$_2$ | ~0 [**900**] | – | ~0 [**900**] | < $10^{-6}$ | Yes | No |
| 561095 | LiHo$_2$Ge$_2$(O$_4$F)$_2$ | ~0 [**900**] | – | ~0 [**900**] | < $10^{-6}$ | Yes | No |
| 8430 | KLiS | ~0 [**900**] | – | ~0 [**900**] | < $10^{-6}$ | Yes | No |
| 22905 | LiCl | N/S | N/S | N/S | ~$10^{-6}$ [from Ref. 8] | Yes | No |

**Table 1:** *Density functional theory molecular dynamics results for machine learning-guided material selection.* We simulate 20 materials as identified by the ML model. For each of the simulated materials, we provide the computed values for diffusivity, conductivity, average Li Bader charge, and the extrapolated RT ionic conductivity values. Here, the diffusivity $\langle D_{ii} \rangle$ represents the average of the diagonal elements of the diffusivity tensor. In parentheses we provide the diffusivities corresponding to one standard deviation above and below the mean when measuring the diffusivity from varying times. We extrapolate these alternate diffusivities to RT to compute expected deivations in the RT conductivity predictions. The second-to-right column gives the ML-based model prediction for superionic likelihood and the rightmost column communicates whether the ML-based model gave a correct prediction for each material. Eight materials have an extrapolated RT ionic conductivity of $10^{-4}$ S/cm or higher, two are below but near this threshold, and ten did not show any Li conduction even at high temperature. Li$_3$InCl$_6$ and LiCl were not simulated ("N/S") because conductivity values could be found in the literature.[8,9] *Li$_2$HIO was only simulated at 400K due to melting at higher temperatures. This extrapolation to RT assumes a diffusion barrier of 0.35 eV, a typical diffusion barrier in fast conducting systems. †CsLi$_2$BS$_3$ could not be simulated at lower temperatures and thus the extrapolated ionic conductivity at RT was taken to be of the same order of magnitude as that of Li$_3$BS$_3$.[24]

ML-chosen candidates, $Li_2GePbS_4$, because of problems with electronic convergence in our simulations.

Our simulations show ten of the simulated candidate materials exhibit significant Li conduction at high temperature. After applying Arrhenius scaling and extrapolating to RT, eight of these materials are predicted to exhibit superionic conductivity (>$10^{-4}$ S/cm) under the simulated vacancy concentrations: $Li_5B_7S_{13}$, $CsLi_2BS_3$, $LiMgB_3(H_9N)_2$, $Li_2B_2S_5$, $Li_3ErCl_6$, $Li_3InCl_6$, $Li_2HIO$, and $LiSO_3F$. Two materials, $Sr_2LiCBr_3N_2$ and $LiErSe_2$, conduct well at high temperature but their extrapolated RT conductivities are below $10^{-4}$ S/cm at RT.

Zhu et al. report[3] that RT superionic conductors will typically exhibit a simulated Li diffusivity of at least 0.01 Å²/ps at 800 K. Of these promising materials that were simulated at 900 K, all have diffusivities significantly above 0.01 Å²/ps. The two marginal conductors are above this limit by a factor of two to three, while the best conductors are above this limit by a factor of ten or more. This provides additional confidence in their fast RT ionic conductivity.

The rightmost column of Table 1 indicates whether the DFT evidence suggests the superionic prediction of the ML model was correct. In total, eight materials are observed to show Li conduction of $10^{-4}$ S/cm or above at RT, two materials are near this threshold, and ten fall far below.

*ii. Discoveries from random selection*

The results of the simulation on the randomly selected materials are listed in Table 2 and plotted in Fig. 2(b). Of the 21 randomly chosen and simulated materials, five structures demonstrated measurable Li conduction at 900 K: $LiSO_3F$, $LiSbO_3$, $Li_6Ho(BO_3)_3$, $Li_2Te_2O_5$, and $RbLi(H_2N)_2$, with Li diffusivities at 900 K of 0.3, 0.1, 0.05, 0.04, and 0.02 Å$^2$/ps respectively. The remaining 16 structures showed no observable Li conduction on the timescale of tens to hundreds of picoseconds. See Supporting Information for computational parameters and simulation times.

The five materials that exhibited diffusion at 900K were simulated again at alternate temperatures. After constructing the conductivity extrapolation, the conductivities in three of these materials ($LiSO_3F$, $Li_6Ho(BO_3)_3$, and $LiSbO_3$) were found to extrapolate to above $10^{-4}$ S/cm at room temperature. The remaining two, however, showed zero diffusion during simulation at 700K. Using typical values for the Li concentration and Bader charge, we estimate that zero Li diffusion after 50ps of simulation at 700K likely corresponds to an ionic conductivity of $10^{-9}$ S/cm or less. Therefore, the RT conductivity of these two materials is predicted to be far below $10^{-4}$ S/cm when the assumed 700K conductivity is factored into the extrapolation.

*iii. Discussion of discovered materials*

Between the ML-guided search and the random search, we identify twelve materials total that are predicted by our DFT-MD simulations to exhibit significant Li ion conduction at RT (and the DFT-MD of Zevgolis et al.[9] for $Li_3InCl_6$). Aside from $Li_3InCl_6$ and $Li_6Ho(BO_3)_3$, none of these materials have reported DFT or experimental conductivity values in the literature to our knowledge. Due to our additional screening steps, these materials also satisfy several other critical criteria beyond fast ionic conductivity to make them useful as solid-state electrolytes in Li-ion batteries: all materials possess DFT-predicted band gaps greater than 1 eV to ensure limited electronic conduction, are free from transition metal elements which may easily reduce in contact with a Li metal anode, exhibit high predicted electrochemical stability against oxidation by cathodes (>4 V vs. Li/Li$^+$), and sit on the convex hull of their phase diagrams to ensure robust structural (phase) stability.

| | MPID (mp-####) | Chemical formula | Computed $\langle D_{ii} \rangle$ (Å²/ps), $T$ (K) | Average Li Bader valence charge | Computed $\langle \sigma_{ii} \rangle$ (mS/cm), T (K) | Extrapolated RT $\langle \sigma_{ii} \rangle$ (mS/cm) | ML-predicted superionic? | Evidence for correct model prediction? |
|---|---|---|---|---|---|---|---|---|
| | 6674 | LiLa$_2$SbO$_6$ | ~0 **[900]** | – | ~0 **[900]** | < 10$^{-6}$ | No | Yes |
| | 8609 | Li$_6$UO$_6$ | ~0 **[900]** | – | ~0 **[900]** | < 10$^{-6}$ | No | Yes |
| | 28567 | LiBiF$_4$ | ~0 **[900]** | – | ~0 **[900]** | < 10$^{-6}$ | No | Yes |
| | 12160 | Li$_6$Ho(BO$_3$)$_3$ | 0.10 (0.12, 0.10) **[1200]**, 0.048 (0.05, 0.044) **[900]** | (+1.0) | 490 (580, 470) **[1200]**, 300 (310, 280) **[900]** | 5.1 (1.7, 3.8) | No | No |
| | 6787 | RbLiB$_4$O$_7$ | ~0 **[900]** | – | ~0 **[900]** | < 10$^{-6}$ | No | Yes |
| | 560072 | Li$_4$Be$_3$As$_3$ClO$_{12}$ | ~0 **[900]** | – | ~0 **[900]** | < 10$^{-6}$ | No | Yes |
| | 7941 | Li$_6$TeO$_6$ | ~0 **[900]** | – | ~0 **[900]** | < 10$^{-6}$ | No | Yes |
| | 13772 | Li$_3$Pr$_2$(BO$_3$)$_3$ | ~0 **[900]** | – | ~0 **[900]** | < 10$^{-6}$ | No | Yes |
| | 8452 | NaLiS | ~0 **[900]** | – | ~0 **[900]** | < 10$^{-6}$ | No | Yes |
| Randomly-guided material selection | 770932 | LiSbO$_3$ | 0.18 (0.19, 0.16) **[1200]**, 0.065 (0.16, 0.078) **[900]** | (+1.0) | 420 (460, 380) **[1200]** 210 (520, 240) **[900]** | 0.67 (520, 5.3) | No | No |
| | 12829 | LiCaGaF$_6$ | ~0 **[900]** | – | ~0 **[900]** | < 10$^{-6}$ | No | Yes |
| | 27811 | Li$_2$Te$_2$O$_5$ | 0.050 (0.14, 0.065) **[1200]**, 0.044 (0.057, 0.042) **[900]**, ~0 **[700]** | (+1.0) | 110 (330, 150) **[1200]** 130 (170, 120) **[900]**, ~0 **[700]** | < 10$^{-6}$ | No | Yes |
| | 8180 | LiNO$_3$ | ~0 **[900]** | – | ~0 **[900]** | < 10$^{-6}$ | No | Yes |
| | 7971 | Ba$_4$Li(SbO$_4$)$_3$ | ~0 **[900]** | – | ~0 **[900]** | < 10$^{-6}$ | No | Yes |
| | 8449 | Rb$_2$Li$_2$SiO$_4$ | ~0 **[900]** | – | ~0 **[900]** | < 10$^{-6}$ | No | Yes |
| | 558045 | NaLi$_2$PO$_4$ | ~0 **[900]** | – | ~0 **[900]** | < 10$^{-6}$ | No | Yes |
| | 562394 | Cs$_4$Li$_2$(Si$_2$O$_5$)$_3$ | ~0 **[900]** | – | ~0 **[900]** | < 10$^{-6}$ | No | Yes |
| | 510073 | RbLi(H$_2$N)$_2$ | 0.088 (0.12, 0.03) **[1100]**, 0.04 (0.036, 0) **[900]**, ~0 **[700]** | (+1.0) | 166 (230, 54) **[1100]**, 84 (76, 0) **[900]**, ~0 **[700]** | < 10$^{-6}$ | No | Yes |
| | 7744 | LiSO$_3$F | 0.86 (0.92, 0.65) **[1200]**, 0.20 (0.22, 0.17) **[900]**, 0.072 (0.08, 0.059) **[700]** | +0.91 | 1300 (1400, 960) **[1100]** 370 (390, 310) **[900]**, 170 (190, 140) **[700]** | 0.082 (0.11, 0.10) | Yes | Yes |
| | 989562 | Cs$_2$LiTlF$_6$ | ~0 **[900]** | – | ~0 **[900]** | < 10$^{-6}$ | No | Yes |
| | 8892 | LiInF$_4$ | ~0 **[900]** | – | ~0 **[900]** | < 10$^{-6}$ | No | Yes |

**Table 2:** *Density functional theory molecular dynamics results for random material selection.* We simulate 21 materials as chosen at random and provide the computed values for diffusivity $\langle D_{ii} \rangle$ (average of diagonal elements in the diffusivity tensor), conductivity, average Li Bader charge, and the extrapolated RT ionic conductivity values. In parentheses, we provide the diffusivities corresponding to one standard deviation above and below the mean when measuring the diffusivity from varying start times. We extrapolate along these alternate diffusivities to RT to compute expected deviations in the RT conductivity predictions. The second-to-right column gives the ML-based model prediction for superionic likelihood and the rightmost column communicates whether the ML-based model gave a correct prediction for each material. The Bader charge on the Li atoms in the control group is taken to be +1.0 for computational efficiency. Additional randomly chosen materials that melted at 900K or whose simulations did not converge were discarded and are not listed here. The success rate of random selection is computed to be 3/21 = 14.3%.

Two of the materials discovered here ($Li_2B_2S_5$ and $Li_5B_7S_{13}$) belong to the Li-B-S system, which appears to be a promising class of materials for realizing fast conducting, electrochemically stable, low mass solid electrolyte materials.[24] The RT ionic conductivity of $Li_5B_7S_{13}$ in particular is predicted to be an exceptional 120 mS/cm, ten times higher than the material $Li_{10}GeP_2S_{12}$ (12 mS/cm),[4] one of the best known Li-ion conductors. $CsLi_2BS_3$ is a Cs-substituted isomorph of $Li_3BS_3$, a fast conductor also in the Li-B-S family. These three phases were initially identified and characterized as phases observed within Li-B-S (and Cs-Li-B-S) glasses.[25–27] The glassy Li-B-S-I system was reported to be an exceptional Li ion conductor in 1981,[28] but to our knowledge these phases within the crystalline Li-B-S space have not been studied as solid electrolytes.

$Li_3ErCl_6$, an Er-substituted analog to the fast-conducting $Li_3InCl_6$, appears to be a promising Li conductor with a predicted RT ionic conductivity of $3\times10^{-4}$ S/cm, although the high atomic mass and low earth abundance of Er makes it a less attractive candidate for battery applications. $Li_3ErCl_6$ was first studied and characterized in 1996 as a potential fast Li ion conductor[29] but to our knowledge the ionic conductivity was not reported. We find another Er-based compound appears to be a marginal RT Li ion conductor, $LiErSe_2$. $LiErSe_2$ is a Li-intercalated two-dimensional layered material that was initially studied in 1987 for its phase transitions and magnetic properties.[30]

We discover three nitride-based fast ion conductor materials with quite complex compositions: $Sr_2LiCBr_3N_2$, $RbLi(H_2N)_2$, and $LiMgB_3(H_9N)_2$. $Sr_2LiCBr_3N_2$ is a solid-state carbodiimide material that was first characterized in 2005;[31] $RbLi(H_2N)_2$ was originally synthesized in 1918[32] and subsequently characterized in 2002;[33] $LiMgB_3(H_9N)_2$ was initially synthesized and studied in 2014 as a candidate solid-state hydrogen storage material.[34] The exotic stoichiometries of these materials underscores the utility of the ML-based model, as it seems unlikely that such complex materials would be discovered by chance to be fast ion conductors. A potential drawback in using these materials with many elements as solid electrolytes is that the electrochemical stability is likely to be low, given the high number of stable interfacial phases that could form from a subset of these elements.

We discover four oxygen-based materials predicted to be fast RT ion conductors: $Li_2HIO$, $LiSO_3F$, $Li_6Ho(BO_3)_3$, and $LiSbO_3$. These materials may be particularly promising candidates for further experimental studies, given that oxide materials can often be synthesized in atmospheric conditions. $LiSbO3$, $LiSO_3F$, and $Li_2HIO$ were first studied in 1954, 1977, and 1994, respectively,[35–37] but to our knowledge have not been studied as a candidate solid electrolyte material. $Li_6Ho(BO_3)_3$ was first characterized[38] in 1977 and experimental work from 1982 reported it to have poor RT ionic conductivity.[39] If the experimental observation is assumed to reflect the "ground truth" of ionic conductivity, this gives $Li_6Ho(BO_3)_3$ the interesting distinction of having been accurately predicted to be a poor ion conductor by the ML model (which was trained on experimental data) but incorrectly predicted to be a good ion conductor by DFT-MD. It is also possible that the bulk ionic conductivity of the material is fast, which is accurately modeled by DFT-MD, but it becomes a poor ion conductor under the experimental conditions of ref. [39], e.g. due to defects, stoichiometric variations, or microstructural effects. The calculations performed here suggest that this material warrants a revisit of the experimental results.

*iv. Model performance*

Quantifying the accuracy of the original ML-based model should be done based on its performance on a randomly-sampled test set, not on a test set consisting only of positive predictions. The 21 randomly chosen materials that were simulated here provide such a set. On this set, the model correctly predicts the DFT-validated material label 19/21 times, yielding an overall predictive accuracy of 90.5%. This value is similar to the 90.0% model accuracy predicted via leave-one-out cross-validation in the original model building process.[7] Only one material ($LiSO_3F$) was predicted to be a superionic conductor by the ML-based model and our DFT simulation confirmed the prediction was correct; this gives a model precision of 1.0 on this particular test set. Of the three materials that DFT-MD found to be RT superionic conductors ($LiSO_3F$, $LiSbO_3$, $Li_6Ho(BO_3)_3$), only one was accurately predicted by the ML model; this gives a model recall of $1/3 = 0.333$ on the test set. Taken together, these values give the model an F1 score of 0.50. For comparison, the baseline F1 score of fast ion conductors on the test set (i.e. the F1 score of completely random guesswork) is simply the fraction of positive examples in the test set, which gives a baseline F1 score of 0.143. Therefore, the F1 score of the ML model is $0.50/0.143 = 3.5x$ higher than the F1 score associated with completely random trial-and-error.

As an alternative means of quantifying the expected improvement in experimental outcomes of Li conductivity measurements in the ML-guided versus random searches, we average the predicted RT ionic conductivities across all randomly chosen materials and all ML-chosen materials. Given that ionic conductivity varies over many orders of magnitude, we average the base-10 logarithm of conductivity. This gives a log-average RT ionic conductivity of the randomly chosen materials of $6.8 \times 10^{-6}$ mS/cm, while the log-average conductivity in the ML chosen materials is $3.1 \times 10^{-4}$ mS/cm, 45 times higher. This comparison assumes the average RT ionic conductivity in the materials with no observed Li motion is $10^{-6}$ mS/cm; as discussed in Supporting Information Section S1, this value likely serves as an upper bound in most cases on the conductivity in these materials, and thus the 45x improvement in conductivity is a lower bound on the true improvement the model offers, e.g. if the average conductivity of the non-conductors is taken to be $10^{-9}$ mS/cm, the improvement in conductivity increases to over 350x.

As an additional performance metric, we compare the likelihood of discovering a RT superionic conductor under the ML-guided screening versus random searching. The random search yields three fast conductors and 18 non-conductors. This yields a baseline superionic probability of 3/21 = 14.3%. To make a fair comparison to the ML-guided case where materials that melted at 900K were not discarded, we do not consider in the statistics those ML-chosen materials that melted at 900K. Two of the eight ML-chosen RT superionic conductors melted at 900K ($LiMgB_3(H_9N)_2$ and $Li_2HIO$). Removing these two materials from the count, the ML model identifies six fast conductors, two marginal conductors, and ten non-conductors. Counting the marginal cases as one half, this gives an ML-guided conductor discovery rate of 7/18 = 38.9%.

Therefore, applying the ML-based model to screen candidate materials before experimentation yields an expected 0.389/0.143 = 2.7x improvement in discovery rate. In Figure 3 we provide a histogram showing the ionic conductivity distribution in the ML-chosen and randomly-chosen materials; the ML model discovers over five times more materials with ionic conductivity above 1 mS/cm, while the number of non-diffusive materials falls from 86% to 56%. We refer the reader to the Supporting Information Section S2 for further discussion.

*v. Comparison to human intuition*

We perform a test of the model's predictive power against human intuition, which is likely to identify fast conducting materials at a higher rate than the background distribution of ion conductors. Several design principles for predicting superionic diffusion have been reported recently,[40–42] but human-guided search efforts may or may not take these principles into account. To understand the precision of human intuition, we polled six Ph.D. students in the Department of Materials Science & Engineering at Stanford University to test their ability to predict superionic conduction in the 21 randomly chosen materials. The students are all actively engaged in research involving ion conductors. The students were allowed to take as much time as necessary to make predictions and could access any structural information about the candidate materials, including space and point groups. After making predictions, we calculated their average precision and recall on the data set; the average precision was 0.25 and the average recall was 0.222, giving an overall average F1 score of 0.235. The baseline F1 score for random guessing is 0.143. The students took approximately one minute to make each individual prediction. For comparison, the ML model made each prediction in approximately one millisecond.

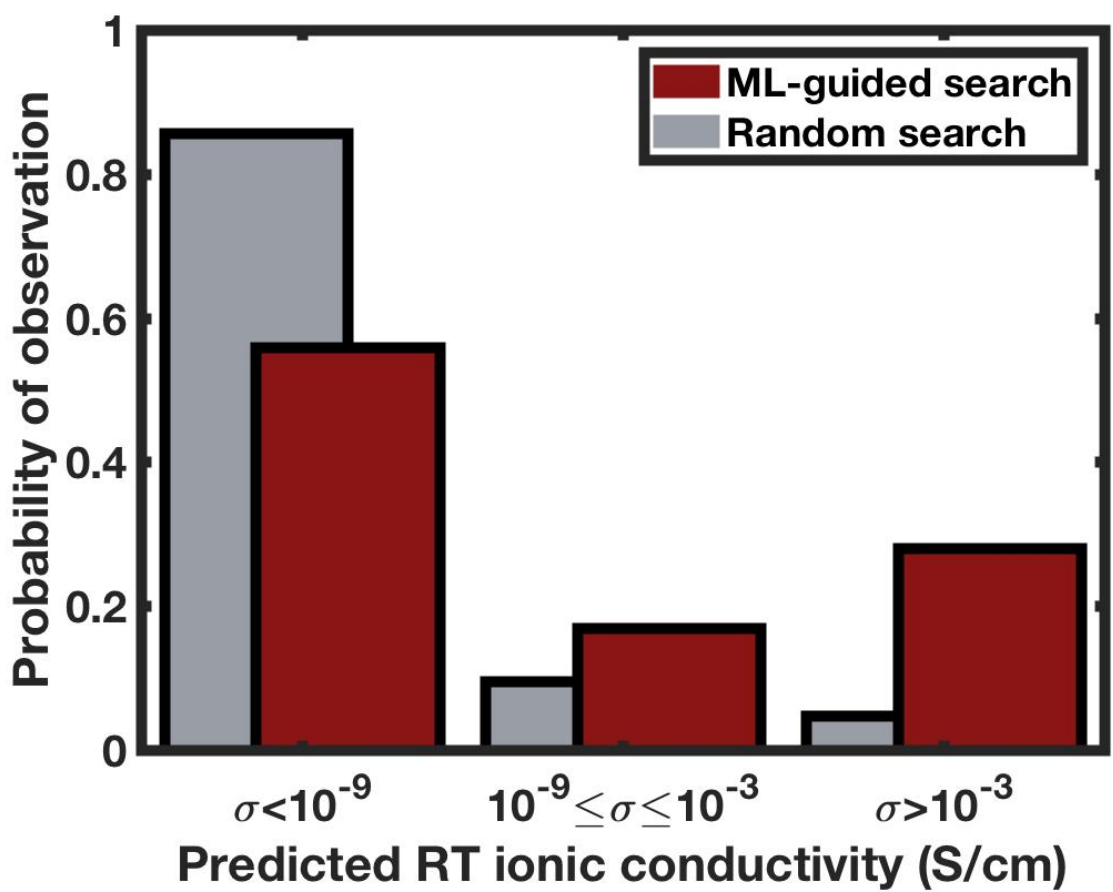

**Figure 3:** *Normalized histogram of predicted room temperature (RT) conductivities for the ML-guided material search versus the random search.* We study 20 materials based on the predictions of our data-driven model, and a control group of 21 chosen at random from the same population. When choosing materials for study based on the predictions of our data-driven model, we observe a significant increase in RT Li conductivity than when simulating materials chosen at random. After removing cases from both populations that melted at 900K, 28% of the ML-selected materials exhibited a predicted RT conductivity of 1 mS/cm or higher with small Li vacancy concentrations, compared to only 5% of materials selected at random. The rate of negative outcomes (no ionic motion) in the simulations during the ML-guided material simulation was 56%, while the randomly chosen materials demonstrated a negative outcome rate of 86%. The log-average of RT predicted conductivity in the ML-guided case is at least 45 times higher if non-conducting cases are assumed to have an RT ionic conductivity $\leq 10^{-6}$ mS/cm ($3.1 \times 10^{-4}$ versus $6.8 \times 10^{-6}$ mS/cm).

In Fig. 4 we compare the performance in speed and F1 score of the Ph.D. students and the ML model. The ML model exhibits more than double the F1 score of the students and is more than three orders of magnitude faster. As a reference, we provide the performance of DFT-MD, which we assume has an F1 score of 1.0 and requires approximately two weeks to make a reliable prediction.

While DFT-MD is taken here to be ground truth, the logistic regression model utilized in this work is trained on experimental measurements, where grain boundaries, contaminants, and other uncharacterized factors may have altered the result. This may mean the model is best suited to guide experimental searches, as the model has a built-in bias towards expected results under experimental conditions. In contrast, our DFT-MD simulates conductivity in single crystal bulk systems with small Li vacancy concentrations. Although DFT-MD is well-suited to make predictions under these conditions, its predictions have potential to deviate from the ground truth for the ML model. For example, the conflicting DFT-MD and experimental reports for RT Li ion conductivity in $Li_6Ho(BO_3)_3$ between this work and ref. [39]; this work predicts RT ion conductivity of approximately 5.1 mS/cm, while ref. [39] reports RT ion conductivity orders of magnitude lower (see Section III, subsection iii). It is possible that the model may recognize that experimental efforts on $Li_6Ho(BO_3)_3$ are likely to yield a poor conductor even though the single crystal conductivity is fast. We look forward to future experimental reports of the ionic conductivities of some of the materials presented here, to both advance the state of the art in ion conductors and to further quantify the performance of the data driven approach.

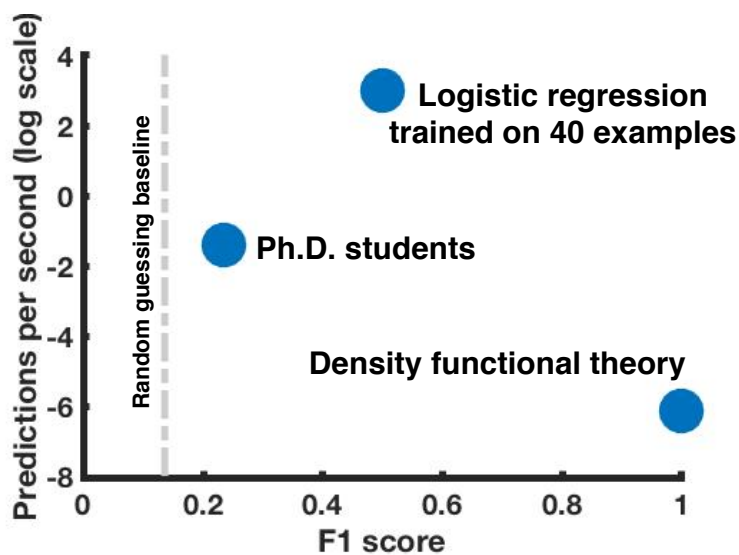

**Figure 4:** *Comparison of human and machine.* Here we plot the predictive accuracy (F1 score) of the ML model and the predictive accuracy of a group of six Ph.D. students working in materials science and electrochemistry. We chose 22 materials at random from a list of 317 highly stable Li containing compounds in the Materials Project database and queried both Ph.D. students and the ML model to predict which materials are superionic Li conductors at room temperature. Validation of superionic conduction is performed with density functional theory molecular dynamics simulation. The Ph.D. students average an F1 score of 0.24 for superionic prediction while the model produces an F1 score of 0.5. For reference, we provide the baseline F1 score expected for random guessing (0.14) as a dashed vertical line. Additionally, the students average a rate of approximately one minute per prediction while the trained ML model makes over 100 predictions per second, a thousand-fold increase in speed. The density functional theory simulation used to validate superionic conduction is a physics-based model that is taken here to represent the ground truth (F1 score of 1.0) but is very slow, with a prediction rate on the scale of weeks. Taken together, these data points highlight the experimental approach that ML-based prediction tools are well-suited to address: making fast predictions more accurately than trial-and-error guessing.

**Conclusions**

Guided by machine learning methods, we discover many new solid materials with predicted superionic Li ion conduction ($\geq 10^{-4}$ S/cm) at room temperature: $Li_5B_7S_{13}$, $Li_2B_2S_5$, $Li_3ErCl_6$, $LiSO_3F$, $Li_3InCl_6$, $Li_2HIO$, $LiMgB_3(H_9N)_2$, and $CsLi_2BS_3$. Two additional materials show marginal RT conduction: $Sr_2LiCBr_3N_2$ and $LiErSe_2$. One of these materials, $Li_5B_7S_{13}$, has a DFT-MD predicted RT Li conductivity (120 mS/cm) an order of magnitude larger than the fastest known Li ion conductors. A search over randomly chosen materials identifies two additional materials with promising predicted RT ionic conductivities: $Li_6Ho(BO_3)_3$, and $LiSbO_3$. In addition to high ionic conductivities, all these materials have high band gaps, high thermodynamic stability, and no transition metals, making them promising candidates for solid-state electrolytes in SSLIBs. These materials represent many exciting new candidates for solid electrolytes in SSLIBs, and we encourage subsequent experimental investigations into the properties of these materials.

Compared to the machine learning-guided search, the control experiment of searching for fast ion conductors through random guesses yields significantly lower quality results. Screening with our ML-based model improves the likelihood of superionic discovery by nearly three times, with a more than 45x improvement in log-average of RT ionic conductivity. Additionally, the model outperforms the intuition of Ph.D. students actively working on ion conductors by more than a factor of two. This is a significant acceleration beyond trial-and-error research and provides confidence in our ML-based approach to materials screening. These results are summarized in Table 3. Such data-driven models are expected to improve with every new data point, and with the data we report here we expect to drive significant improvements to ML-based models for Li conduction.

| Model | Precision | Recall | F1 Score | Fast Li ion conducting materials identified (in this work) |
|---|---|---|---|---|
| *Machine learning (logistic regression on 40 examples)* | 1.0 | 0.33 | 0.5 | $Li_5B_7S_{13}$, $Li_2B_2S_5$, $Li_3ErCl_6$, $Li_2HIO$, $LiSO_3F$, $Li_3InCl_6$, $CsLi_2BS_3$, $LiMgB_3(H_9N)_2$, $Sr_2LiCBr_3N_2$ (marginal), $LiErSe_2$ (marginal), |
| *Random selection* | 0.14 | 0.14 | 0.14 | $Li_6Ho(BO_3)_3$, $LiSbO_3$, $LiSO_3F$ |
| *Ph.D. student screening* | 0.25 | 0.22 | 0.24 | – |

**Table 3: *Summary of model performance*.** Here we provide the performance metrics (precision, recall, F1 score) and list of discovered fast Li ion conducting materials for the three models explored in this work. The results of the Ph.D. student screening represent the average performance of six students on the same test set of randomly-chosen materials as used to compute the "random selection" statistics. The machine learning-based approach to predicting ion conductivity from a small data set of 40 materials significantly outperforms both random chance and the average polled Ph.D. student. This highlights the promise of applying machine learning-based approaches to materials screening before performing computationally expensive simulation techniques like DFT-MD or time consuming experimental tests.

The improvement over random guessing provided by our ML-based model for predicting Li ion conductivity underscores the importance of thorough data reporting, centralized data collection, and careful data analysis for materials. Although significant improvements have been made in the last several years thanks to incentives provided by the Materials Genome Initiative,[43] there are still many materials properties, especially experimentally expensive properties like ionic conductivity, without any comprehensive data repository. This work demonstrates that learning on even small sets of materials data (40 samples) can offer a significant advantage in screening efforts. To that end, we encourage efforts to continue centralizing and learning on diverse types of data on materials.

**Acknowledgements**

This work is funded in part by a seed grant from the TomKat Center for Sustainable Energy at Stanford University and in part by Toyota Research Institute through the Accelerated Materials Design and Discovery Program. The authors acknowledge helpful contributions from Daniel Rehn and constructive conversations with Profs. Nicole Adelstein, Geoffroy Hautier, and Mauro Pasta, Dr. Brandon Wood, and Leonid Kahle.

**Supporting Information Description**

In the Supporting Information, we provide a table of the computational parameters for the DFT simulations, calculations for the ionic conductivity upper bound in the non-diffusive materials,

and discussion on the impact on the model performance metrics of the removal of the randomly-

chosen

# SUPPORTING INFORMATION
Machine learning-assisted discovery of many new solid Li-ion conducting materials

| | MPID (mp-####) | Chemical formula | Atoms in unit cell | Li vacancy concentration | Simulation temperatures (K) | Simulation time (ps) | VASP PAW-PBE pseudopotentials |
|---|---|---|---|---|---|---|---|
| ML-guided material selection | 554076 | $BaLiBS_3$ | 95 | 6.3% | 900 | 80.4 | Ba_sv, Li_sv, B, S |
| | 532413 | $Li_5B_7S_{13}$ | 99 | 5% | 900, 700, 400 | 33.3, 211.0, 113.8 | Li_sv, B, S |
| | 569782 | $Sr_2LiCBr_3N_2$ | 71 | 12.5% | 1200, 900 | 126.0 | Sr_sv, Li_sv, C, Br, N |
| | 558219 | $SrLi(BS_2)_3$ | 87 | 12.5% | 900 | 86.5 | Sr_sv, Li_sv, B, S |
| | 15797 | $LiErSe_2$ | 47 | 8.3% | 1200, 900 | 28.4, 59.0 | Li_sv, Er_3, Se |
| | 29410 | $Li_2B_2S_5$ | 71 | 6.3% | 900, 700, 400 | 115.7, 49.4, 54.0 | Li_sv, B, S |
| | 676361 | $Li_3ErCl_6$ | 59 | 5.6% | 900, 700 | 38.7 | Li_sv, Er_3, Cl |
| | 643069 | $Li_2HIO$ | 79 | 3.1% | 400 | 169.6 | Li_sv, H, I, O |
| | 7744 | $LiSO_3F$ | 95 | 6.3% | 1100, 900, 700 | 13.1, 103.7, 89.8 | Li_sv, S, O, F |
| | 34477 | $LiSmS_2$ | 63 | 6.3% | 900 | 130.7 | Li_sv, Sm_3, S |
| | 559238 | $CsLi_2BS_3$ | 111 | 3.1% | 900 | 17.2 | Cs_sv, Li_sv, B, S |
| | 866665 | $LiMgB_3(H_9N)_2$ | 149 | 16.7% | 700, 500 | 77.3 | Li_sv, Mg_pv, B, H, N |
| | 8751 | RbLiS | 53 | 5.6% | 900 | 69.3 | Rb_sv, Li_sv, S |
| | 15789 | $LiDyS_2$ | 105 | 12.5% | 900 | 76.8 | Li_sv, Dy_3, S |
| | 15790 | $LiHoS_2$ | 71 | 5.6% | 900 | 78.9 | Li_sv, Ho_3, S |
| | 15791 | $LiErS_2$ | 71 | 5.6% | 900 | 26.6 | Li_sv, Er_3, S |
| | 561095 | $LiHo_2Ge_2(O_4F)_2$ | 127 | 12.5% | 900 | 67.6 | Li_sv, Ho_3, Ge_d, O, F |
| | 8430 | KLiS | 53 | 5.6% | 900 | 50.1 | K_sv, Li_sv, S |
| Randomly-guided material selection | 6674 | $LiLa_2SbO_6$ | 79 | 12.5% | 900 | 118.5 | Li_sv, La, Sb, O |
| | 8609 | $Li_6UO_6$ | 311 | 0.7% | 900 | 21.2 | Li_sv, U, O |
| | 28567 | $LiBiF_4$ | 95 | 6.3% | 900 | 107.7 | Li_sv, Bi, F |
| | 12160 | $Li_6Ho(BO_3)_3$ | 151 | 2.1% | 1200, 900 | 38.7, 60.0 | Li_sv, Ho_3, B, O |
| | 6787 | $RbLiB_4O_7$ | 103 | 12.5% | 900 | 89.1 | Rb_sv, Li_sv, B, O |
| | 560072 | $Li_4Be_3As_3ClO_{12}$ | 91 | 6.3% | 900 | 89.3 | Li_sv, Be_sv, As, Cl, O |
| | 7941 | $Li_6TeO_6$ | 311 | 0.7% | 900 | 30.4 | Li_sv, Te, O |
| | 13772 | $Li_3Pr_2(BO_3)_3$ | 135 | 4.2% | 900 | 43.9 | Li_sv, Pr_3, B, O |
| | 8452 | NaLiS | 53 | 5.6% | 900 | 191.8 | Na_pv, Li_sv, S |
| | 770932 | $LiSbO_3$ | 79 | 6.3% | 1200, 900 | 97.1, 86.3 | Li_sv, Sb, O |
| | 12829 | $LiCaGaF_6$ | 71 | 12.5% | 900 | 100.7 | Li_sv, Ca_sv, Ga_d, F |
| | 27811 | $Li_2Te_2O_5$ | 71 | 6.3% | 1200, 900, 700 | 19.1, 147.6, 27.7 | Li_sv, Te, O |
| | 8180 | $LiNO_3$ | 79 | 6.3% | 900 | 50.3 | Li_sv, N, O |
| | 7971 | $Ba_4Li(SbO_4)_3$ | 79 | 25% | 900 | 84.8 | Ba_sv, Li_sv, Sb, O |

| | | | | | | |
|---|---|---|---|---|---|---|
| 8449 | $Rb_2Li_2SiO_4$ | 71 | 6.3% | 900 | 94.9 | Rb_sv, Li_sv, Si, O |
| 558045 | $NaLi_2PO_4$ | 63 | 6.3% | 900 | 135.3 | Na_pv, Li_sv, P, O |
| 562394 | $Cs_4Li_2(Si_2O_5)_3$ | 107 | 12.5% | 900 | 31.7 | Cs_sv, Li_sv, Si, O |
| 510073 | $RbLi(H_2N)_2$ | 95 | 8.3% | 1100, 900, 700 | 23.1, 122.6, 28.3 | Rb_sv, Li_sv, H, N |
| 7744 | $LiSO_3F$ | 95 | 6.3% | 1100, 900, 700 | 13.1, 103.7, 89.8 | Li_sv, S, O, F |
| 989562 | $Cs_2LiTlF_6$ | 79 | 12.5% | 900 | 42.3 | Cs_sv, Li_sv, Tl_d, F |
| 8892 | $LiInF_4$ | 95 | 6.3% | 900 | 65.4 | Li_sv, In_d, F |

**Table S1:** *Computational parameters.* We perform DFT-MD calculations using the above parameters.

## Section S1: Room temperature conductivity of non-diffusive materials

When no Li motion is observed during the course of the DFT-MD simulation, there is no information on which to base an ionic conductivity estimate. However, it is possible to bound this estimate by considering what the ionic conductivity would be if one inter-site Li hop was observed immediately at the end of the simulation. This corresponds to an upper bound on conductivity, as many more picoseconds of simulation time may be required before this hop is actually observed.

We first apply the mean squared displacement approach to computing diffusivity formula, eq. (2). We assume that, on average, approximately 100ps of simulation time passes before a material is considered to be non-conductive at 900K. We also assume that there are 21 Li atoms in the unit cell and the inter-site hopping distance for the one diffusing Li atom is 3.8 Å, the average number of Li atoms and average Li-Li distance across all the materials simulated in this work. Introducing one Li vacancy gives 20 Li atoms in the computational cell. In terms of the variables in eq. (2), $\Delta t = 100$ps and $\langle \Delta r^2 \rangle = (3.8\text{Å})^2/(20 \text{ Li atoms}) = 0.72 \text{ Å}^2$, which gives an upper bound Li diffusivity at 900K of $D = \frac{1}{6}\left(\frac{0.72 \text{ Å}^2}{100 \text{ ps}}\right) = 0.0012 \text{ Å}^2/\text{ps}$. We convert to ionic conductivity using the Einstein relation assuming the Li Bader charge $q = +1.0$, and the Li concentration $n = 0.013 \text{ Å}^{-3}$ (the average Li concentration across all materials simulated here). This gives an upper bound 900K ionic conductivity of $\sigma(900K) = (0.0012 \text{ Å}^2/\text{ps})(0.013 \text{ Å}^{-3})(1.6 \times 10^{-19} \text{ C})^2/(k_B \times 900K) = 3$ mS/cm. Typically, poorly conducting materials will exhibit diffusion barriers of approximately $E_a = 0.5$ eV or higher. Taking this to be a typical barrier, we extrapolate the estimated conductivity at 900K down to 293K assuming Arrhenius scaling: $\sigma(T) = \sigma_0 e^{-E_a/k_B T}$. This gives an upper bound RT conductivity under these conditions of approximately $10^{-6}$ mS/cm.

## Section S2: Impact of removal of melting candidate materials on model performance metrics

When computing the improvement in superionic discovery rate offered by the model over random guesswork, we remove cases that melted at 900 K from both data sets. In the ML-guided case, we discover that the two materials that melted at 900 K are in fact predicted by DFT to be fast RT conductors. We discover this by simulating these materials at increasingly lower temperatures until no melting is observed and then extrapolating to RT. This procedure was not followed for the randomly-selected cases.

If we do not remove the melting cases from the ML-guided materials discovery statistics, the rate of discovery for RT superionic conductors becomes 9/20 = 45.0%. If we assume the "worst case scenario" that the two randomly-chosen cases that melted at 900 K are in fact fast Li-ion conductors at RT, this would increase the background probability of superionic conductors from 3/21 = 14.3% to 5/23 = 21.7%. The F1 score of completely random guessing then correspondingly increases to 0.217. Under the "best case scenario" that they are not truly fast Li-ion conductors, the background probability falls to 3/23 = 13.0% and the corresponding F1 socre is 0.130. This potential variation in the discovery rates results in a potential variation in the expected improvement over random offered by the model. In the best case scenario, the expected improvement in discovery rate improves to 0.45/0.130 = 3.5x. Under the worst case scenario, the expected improvement falls to 0.45/0.217 = 2.1x.